\documentclass[prl,superscriptaddress,showpacs,floatfix]{revtex4-2}
\usepackage{graphicx}
\usepackage{dcolumn}
\usepackage{braket}
\usepackage[normalem]{ulem}
\usepackage{bm}
\usepackage{amsfonts} 
\usepackage{amsmath} 
\usepackage[bookmarksnumbered,pdfpagelabels=true,plainpages=false,colorlinks=true,linkcolor=blue,citecolor=blue,urlcolor=blue]{hyperref}
\usepackage[bookmarksnumbered,pdfpagelabels=true,plainpages=false,colorlinks=true,linkcolor=blue,citecolor=blue,urlcolor=blue]{hyperref}

\def \usach {Departamento de F\'isica, Universidad de Santiago de Chile, 9170124, Santiago, Chile.}
\def \cedenna {Centro  de Nanociencia y Nanotecnología CEDENNA, Avda. Ecuador 3493, Santiago, Chile.}

\begin{document}

\title{On the Electronic Contribution to Crystalline Diffraction Patterns}

\author{Sebastian Allende}
\email{sebastian.allende@usach.cl}
\affiliation{\usach}
\affiliation{\cedenna}

\author{David Galvez-Poblete}
\affiliation{\usach}
\affiliation{\cedenna}

\begin{abstract}
We introduce the electronic structure factor as a 
phase-sensitive contribution to diffraction that directly encodes the 
properties of the occupied-band wave functions. In the one-dimensional SSH 
model, $F_{\mathrm{cond}}$ is governed by the relative sublattice phase, 
which integrates to the Zak phase. This provides a clear diffraction-based 
criterion to distinguish trivial and topological regimes in the absence of 
any structural change. Beyond the SSH limit, the same Bloch-based 
construction naturally accounts for commensurate and incommensurate 
magnetic satellites in antiferromagnets, reproducing the additional peaks 
at $q=G\pm Q$ observed in NiO, MnO, chromium, and cuprates. These results 
demonstrate that diffraction can probe electronic topology and magnetic 
ordering on equal footing, opening a route to phase-sensitive structural 
characterization of correlated electron systems.
\end{abstract}

\maketitle

\section{Introduction}

In recent decades, the discovery of novel electronic phases in solids 
has revealed a wealth of phenomena that cannot be understood solely 
in terms of conventional structural mechanisms. Examples include the 
metal--insulator transition driven by electron correlations 
\cite{Imada1998}, the emergence of topological phases such as 
topological insulators\cite{Hasan2010,Qi2011,Fu2007,Roy2009} and Weyl \cite{Armitage2018,Xu2015,Yan2017}, and the appearance of magnetic order 
such as commensurate and incommensurate antiferromagnetism\cite{Fawcett1988,Shibatani1969,deOliveira2007,Bisti2025}. 
A common feature of these systems is that their essential signatures 
are predominantly electronic and may occur without any visible 
structural distortion.

Detecting such phases is therefore challenging, since traditional 
diffraction probes are designed to reveal the nuclear lattice but 
remain blind to purely electronic reorganizations. Instead, their 
identification typically relies on indirect observables: changes in 
transport coefficients \cite{Konig2007,Huang2015}, edge or surface states 
\cite{Hsieh2008,Chen2009}, and advanced spectroscopies such as ARPES or 
resonant x-ray scattering 
\cite{Damascelli2003,Lv2015,Ament2011}. 
While powerful, these approaches often require complex experimental 
setups, special geometries, or prior knowledge of the expected order.

Scattering and diffraction techniques have been extensively employed for more than a century as powerful methods to probe the atomic arrangement and electronic charge density of materials\cite{Kittel1967}. The fundamental principle relies on the interaction of an incident beam—such as electrons, neutrons, or X-rays with the nuclei and electrons in the sample. The resulting scattered waves interfere and generate a diffraction pattern whose intensity distribution is determined by the structure factor $S(q)$, directly related to the spatial distribution of atoms and electronic density. Analysis of these patterns provides essential information about the microscopic nature of the material. While these methods have traditionally been applied to elucidate crystallographic structures with high precision\cite{Giacovazzo2011}, the underlying physics is considerably richer: scattering experiments also encode information on electronic correlations, collective excitations, and even topological aspects of matter, offering a much broader window into the fundamental properties of condensed systems.

In this work, we propose an extension of the diffraction formalism 
that explicitly incorporates the contribution of conduction electrons 
to the structure factor. This formulation provides a unified framework 
to access electronic reorganizations directly through diffraction, even 
when the crystalline lattice remains unchanged. We illustrate the 
approach in three representative cases: the metal--insulator transition, 
topological electronic phases, and commensurate and incommensurate 
antiferromagnetic order.

\section{The model}

The diffraction intensity, $I(\vec{q})$, for a wave incident on a crystalline solid is given by the square of the modulus of the structure factor ($F(\vec{q})$)\cite{Marder2010}:
\begin{equation}
I(\vec{q}) \propto |F(\vec{q})|^2
\end{equation}

The momentum transfer vector is $\vec{q} = \vec{k}_{\text{dif}} - \vec{k}_0$, where $\vec{k}_0$ is the incident wavevector and $\vec{k}_{\text{dif}}$ is the diffracted wavevector, with  $|\vec{k}_0| = |\vec{k}_{\text{dif}}| = 2\pi/\lambda$ for elastic processes.

The structure factor $F(\vec{q})$ is the Fourier transform of the total electron density of the material:
\begin{equation}
F(\vec{q}) = \int \rho_{\text{el}}(\vec{r}) \, e^{i \vec{q} \cdot \vec{r}} \, d^3r
\end{equation}

where $\rho_{\text{el}}(\vec{r})$ is the electron density bound to the \textit{N} \text{atoms} in the cell, centered at positions $\vec{r}_j$, with local profiles. Then $\rho_{\text{el}}(\vec{r})$ is written as (Kittel):

  \begin{equation}
  \rho_{\text{el}}(\vec{r}) = \sum_j \rho_j(\vec{r} - \vec{r}_j)
  \end{equation}

However, this representation is not unique, and this is where we come in. Within the framework of density functional theory (DFT)\cite{Hohenberg1964,Kohn1965} or pseudopotentials \cite{Pickett1989}, it is common to separate the electron density into two contributions, one corresponding to the electrons in the ion/core through an effective potential and the other considering the valence electrons actively. In other words, we will separate the electron density into a local part (ion/core) and an extended part, that is:

\begin{equation}
\rho_{\text{el}}(\vec{r}) = \rho_{\text{ion}}(\vec{r}) + \rho_{\text{cond}}(\vec{r})
\end{equation}

where  $\rho_{\text{ion}}(\mathbf r)=\sum_j \rho^{\text{core}}_j(\mathbf r-\mathbf r_j)$ refers to the ionic (or core) electron density, which is commonly treated as a static effective potential. It excludes the valence electrons and is assumed to remain unchanged across phase transitions, since it corresponds to electrons in the inner atomic shells. In contrast, $\rho_{\text{cond}}(\vec{r})$ represents the electronic density associated with the outermost valence and conduction band states, and it depends sensitively on how the available electronic states are occupied near the Fermi level. his contribution evolves with doping, temperature, and external fields, and therefore directly reflects the reorganizations of the electronic structure. That is:

\begin{equation}
\rho_{\text{cond}}(\vec{r}) = \sum_{n\vec{k},\sigma} |\psi_{n\vec{k},\sigma}(\vec{r})|^2 \, f(E_{n\vec{k},\sigma} - \mu)
\label{rhocond}
\end{equation}

where $f(E_{n\vec{k},\sigma} - \mu)$ is the Fermi-Dirac function, $\mu$ is the chemical potential,

\begin{equation}
\psi_{n\vec{k},\sigma}(\vec{r}) = u_{n\vec{k}\sigma}(\vec{r}) e^{i\vec{k} \cdot \vec{r}}
\end{equation}

correspond to Bloch states in a three-dimensional crystal with energy $E_{n\vec{k},\sigma}$. In Eq. \ref{rhocond} the summation also runs over the spin configurations $\sigma$. For simplicity, this contribution will be neglected in the following expressions, since the results can be straightforwardly extended to include spin. Therefore, equation \ref{rhocond} is given by: 

\begin{equation}
\rho_{\text{cond}}(\vec{r}) = \sum_{n\vec{k}} |u_{n\vec{k}}(\vec{r})|^2 \, f(E_{n\vec{k}} - \mu).
\end{equation}

This representation of the electronic density is well established in 
density functional theory (DFT), band theory, and tight-binding 
models~\cite{Slater1954,Papaconstantopoulos2015}. 
The separation between $\rho_{\text{ion}}$ and $\rho_{\text{cond}}$ 
distinguishes localized core electrons from the extended band states 
that govern transport. 
The spatial structure of $\rho_{\text{cond}}(\vec{r})$ depends on the 
Fermi level, temperature, doping, or external fields, and thus evolves 
across different electronic regimes. 
In what follows, we show that promoting this contribution to an explicit 
structure factor $F_{\mathrm{cond}}$ provides a natural framework to 
capture how electronic reorganizations manifest in diffraction, even in 
the absence of structural changes. This dependence is key to explaining the differences in diffraction intensity in metallic, semiconductor, or insulating phases.

With this, let us calculate the structure factor in the Laue condition $\vec{q}=\vec{G}$, where $\vec{G}$ is a vector of the reciprocal lattice. Then we have:

\begin{equation}
F(\vec{q}) = F_{\text{ion}}(\vec{q})+ F_{\text{cond}}(\vec{q})
\end{equation}

where, 

\begin{equation}
F_{\text{ion}}(\vec{q}) = \sum_j f_j(\vec{q}) e^{i\vec{q}\cdot\vec{r}_j} 
\end{equation}

with 

\begin{equation}
  f_j(\vec{q}) = \int \rho^{\text{core}}_j(\vec{r}) \, e^{i\vec{q}\cdot\vec{r}} \, d^3r,
\end{equation}

and

\begin{align}
F_{\text{cond}}(\vec{q}) &= \sum_{n\vec{k}} f(E_{n\vec{k}} - \mu) \int |u_{n\vec{k}}(\vec{r})|^2 \, e^{i\vec{q} \cdot \vec{r}} \, d^3r \nonumber \\
&= \sum_{n\vec{k}} f(E_{n\vec{k}} - \mu) \, S_{n\vec{k}}(\vec{q}).
\label{fcond}
\end{align}

Here

\begin{equation}
    S_{n\vec{k}}(\vec{q})=\int_{\text{cell}} |u_{n\vec{k}}(\vec{r})|^2 \, e^{i\vec{q} \cdot \vec{r}} \, d^3r.
\end{equation}

Therefore, We define the band-electron contribution to the structure factor, $F_{\text{cond}}(\vec{q})$, arising from the extended valence/conduction states. Unlike the ionic channel $F_{\text{ion}}$, his term can retain explicit information about the relative phase of the Bloch wave functions.

\section{Results and discussion}

Using this structure factor for the conduction electron density, we now consider three  cases: The first case concerns detecting the system’s topological phase. To this end, we use the Su–Schrieffer–Heeger (SSH) model\cite{Su1979,Su1980,Asbth2016} as an example. Consider a one-dimensional chain with two sublattices per unit cell: $A$ at $x=R$ and $B$ at $x=R+a/2$. Let $w_A(x)$ and $w_B(x)$ denote the (real, normalized) Wannier functions centered at $A$ and $B$, respectively. For a Bloch state in the occupied band we write \cite{Atala2013} 
\begin{equation}
  \psi_k(x)=\frac{\phi_A(k)}{\sqrt{N}}\sum_R e^{ikR} w_A(x-R)
          + \frac{\phi_B(k)}{\sqrt{N}}\sum_R e^{ik(R+a/2)} w_B\big(x-R-a/2\big),
  \label{eq:BlochWannier}
\end{equation}
where $(\phi_A,\phi_B)$ is the spinor in the sublattice basis. We then have that equation~(\ref{fcond}) is given by:

\begin{align}
  F_{\mathrm{cond}}(q)
  &= \left[\int_{\mathrm{BZ}}\frac{dk}{2\pi/a}\, f\,|\phi_A(k)|^2\right] S_A(q)S_0(q)
   + \left[\int_{\mathrm{BZ}}\frac{dk}{2\pi/a}\, f\,|\phi_B(k)|^2\right] S_B(q)\,e^{iqa/2}S_0(q)
   \nonumber\\
  &\quad + \left[2\,\Re\int_{\mathrm{BZ}}\frac{dk}{2\pi/a}\, f\,\phi_A(k)\phi_B^*(k)e^{ika/2}\right]
       S_{\mathrm{bond}}(q)\,e^{iqa/2}S_0(q),
  \label{Fcond_general}
  \end{align}

where $f\equiv f(E(k)-\mu,T)$ is the Fermi–Dirac function, 

\begin{align}
  S_A(q) &= \int dx\,|w_A(x)|^2\,e^{iqx}, \qquad
  S_B(q) = \int dx\,|w_B(x)|^2\,e^{iqx}, \\
  S_{\mathrm{bond}}(q) &= \int dx\,w_A(x)\,w_B^*(x-a/2)\,e^{iqx},
\end{align}

and $S_0(q)=\sum_{R}e^{iqR}=\sum_{n=0}^{N-1}e^{iqna}$ is the lattice factor. In the limit $N\to\infty$, $S_0(q)\to\frac{2\pi}{a}\sum_{m\in\mathbb Z}\delta(q-G_m)$, with $ G_m=\frac{2\pi}{a}m$, which enforces the Laue condition. The equation~(\ref{Fcond_general}) was obtained by performing the translations
$x\mapsto x+R$ for $S_A(q)$, $x\mapsto x+R+a/2$ for $S_B(q)$, and $x\mapsto x+R$ for $S_{bond}(q)$. If we now introduce the averages over the first Brillouin zone as:
\begin{equation}
  C_A=\int\frac{dk}{2\pi/a} f\,|\phi_A|^2,\quad
  C_B=\int\frac{dk}{2\pi/a} f\,|\phi_B|^2,\quad
  C_{\mathrm{bond}}=\int\frac{dk}{2\pi/a} f\,\phi_A\phi_B^* e^{ika/2},
\end{equation}
we finally obtain

\begin{equation}
  F_{\mathrm{cond}}(q)=
  \Big[ C_A S_A(q) + C_B S_B(q)e^{iqa/2} + 2\,\Re C_{\mathrm{bond}} \, S_{\mathrm{bond}}(q)\,e^{iqa/2} \Big]\;
  S_0(q).
  \label{eFcond_compactoe}
\end{equation}

Since $S_0(q)$ is a sum of deltas, the intensity is only nonzero at Bragg points $q=G_m$, Laue condition. Therefore $e^{iGa/2}=(-1)^m$ and  equation~(\ref{eFcond_compactoe}) becomes:

\begin{equation}
  F_{\mathrm{cond}}(G)=
  \Big[ C_A S_A(G) + (-1)^m C_B S_B(G) + (-1)^m 2\,\Re C_{\mathrm{bond}} \, S_{\mathrm{bond}}(G) \Big]\;
  S_0(q).
  \label{eq:Fcond_Bragg}
\end{equation}

Then, if we assume $S_A(G)\approx S_B(G)=S(G)$, the site contribution of $F_{\mathrm{cond}}(G)$ reduces to
$F_{\mathrm{site}}(G)=S(G)\,\big[ C_A + (-1)^m C_B \big]$.
As can be seen from $F_{\mathrm{site}}$, this term does not carry the system’s topological information.
Next we show how the term $F_{\mathrm{topo}}=F_{\mathrm{cond}}-F_{\mathrm{site}}$ represents the part that encodes the system’s topological information; specifically, we will show how the factor $\phi_A\phi_B^*$ transports the relative phase through this product.
To simplify our calculations, consider the ideal SSH case with chiral symmetry ($d_z=0$).
Then we have $\phi_A(k) = -e^{i\alpha(k)}/\sqrt{2}$ and $\phi_B(k) = e^{i\beta(k)}/\sqrt{2}$.
It then follows that the term $F_{\mathrm{topo}}$ becomes:

\begin{equation}
  F_{\mathrm{topo}}(G)\;=\;
   -\,(-1)^m\, \Re\,\mathcal{I}_{\mathrm{topo}}\;\, S_{\mathrm{bond}}(G)\;S_0(q)
  \label{eq:main_result}
\end{equation}

where $\Re\,\mathcal{I}_{\mathrm{topo}}=-2\, \Re C_{\mathrm{bond}}$ and $\mathcal{I}_{\mathrm{topo}}$ is given by:

\begin{equation}
  \mathcal{I}_{\mathrm{topo}}
  = \int_{\mathrm{BZ}}\frac{dk}{2\pi/a}\, f\big(E(k)-\mu\big)\; e^{-i\Phi(k)}\,e^{ika/2}.
  \label{eq:Itopo}
\end{equation}

Here $\Phi(k)$ is the relative phase between $\phi_A(k)$ and $\phi_B(k)$, that is, $\Phi(k)=\alpha(k)-\beta(k)$, which is related to the Berry connection via $\mathcal{A}(k)=\tfrac{1}{2}\,\partial_k \Phi(k)$ (see the appendix for the derivation). Then $\Phi(k)$ naturally connects to the Zak phase as 
$\gamma = \int_{\mathrm{BZ}} \mathcal{A}(k)\,dk = \tfrac{1}{2}\int_{\mathrm{BZ}} \partial_k \Phi(k)\,dk = \tfrac{1}{2}\,\Delta \Phi \equiv \pi\,w \ \ (\mathrm{mod}\ 2\pi)$,
where $w$ is the winding number of $\Phi(k)$ as $k$ traverses the Brillouin zone. The key ingredient is the relative phase $\Phi(k)$, which enters only through the conduction channel $F_{\mathrm{cond}}$ but not through the ionic contribution. This makes the electronic part intrinsically phase-sensitive, allowing diffraction to access the Zak phase. This phase-sensitivity is the novel aspect of our formulation: $F_{\mathrm{cond}}$ is not simply an additive correction to the ionic form factor, but a distinct channel that directly encodes the relative phase of the occupied-band spinors. Now, if the experiment is tuned so that \(F_{\mathrm{ion}}(\mathbf G)\approx 0\) and a reflection with odd intracell phase (odd \(m\)) is chosen, the intensity is dominated by the electronic channel, which in this case is topological:
$I( G)\simeq |F_{\mathrm{cond}}(G)|^2=\lvert F_{\mathrm{topo}}( G)\rvert^{2} \propto \mathcal{I}^2_{\mathrm{topo}}$.
This, in turn, means that the intensity $I(\mathbf G)$ will be affected by whether the system is in its trivial phase ($w=0$) or its topological phase ($w=1$), depending on the value of the ratio $t_1/t_2$ that sets $\Phi(k)=\arg\big(d_x+ i d_y\big)$, where $d_x(k)=t_1+t_2\cos(ka)$ and $d_y(k)=t_2\sin(ka)$.

\begin{figure}[bth!]
    \centering
    \includegraphics[width=0.6\linewidth]{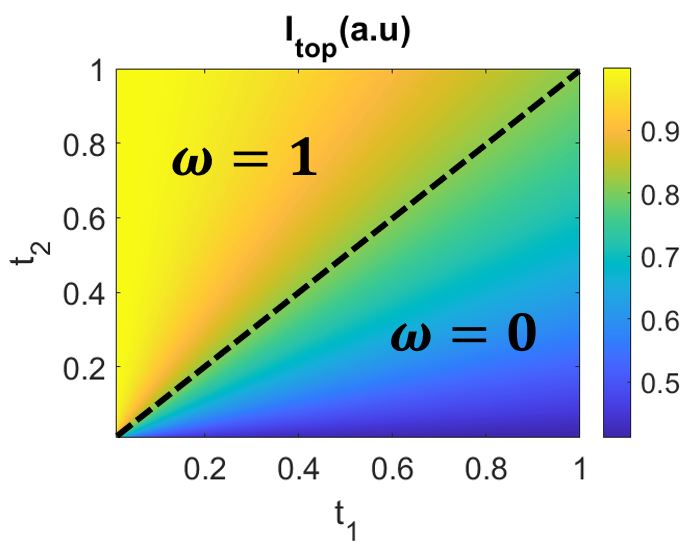}
    \caption{Normalized intensity of the topological term with respect to $S^2_{bond}(G)$. The dashed line indicates the metallic phase $(t_1 = t_2)$ in the SSH model. The upper diagonal region corresponds to the topological phase, characterized by winding number $\omega =1$, while the lower region corresponds to the trivial phase with $\omega =0$. }
    \label{fig:placeholder}
\end{figure}

We introduce an electronic structure factor \(F_{\mathrm{cond}}\) that renders diffraction intensities phase sensitive. Under reflections with \(e^{iGa/2}=-1\) and \(F_{\mathrm{ion}}(G)\approx 0\), the signal reduces to \(I(G)\propto\lvert F_{\mathrm{topo}}(G)\rvert^{2}\), governed by the relative sublattice phase that encodes the Zak phase \((\pi w)\) in SSH. This enables diffraction-based discrimination between trivial and topological phases, field tunable and independent of structural changes.

The second case is the appearance of commensurate or incommensurate magnetic 
Bragg satellites in antiferromagnets (AFM), such as NiO or MnO for the 
commensurate case\cite{Shull1949,Roth1958}, and chromium for the incommensurate one\cite{Fawcett1988}. We start from 
the antiferromagnetic Bloch,
\begin{equation}
\psi_k(r) = \alpha_k e^{ikr}u_A(r) + \beta_k e^{i(k+Q)r}u_B(r),
\end{equation}
where $u_{A,B}(r)$ are periodic over the crystallographic cell of size $a$. In an antiferromagnetic material, the effective potential due to the spin 
modulation has a doubled periodicity of $2a$, hence the ordering vector is 
$Q=\pi/a$, which folds the Brillouin zone, even though the underlying 
crystal lattice still has period $a$. As a consequence, electronic Bloch states with 
momenta $k$ and $k+Q$ become coupled. In the commensurate case, 
$Q=\pi/a$ corresponds to the ordering wave vector of the antiferromagnetic 
modulation. For the incommensurate case, the ordering vector takes the form $Q = \frac{\pi}{a} + \delta$, where $\delta$ measures the deviation from perfect commensurability.  

The conduction electron density is then written as $\rho_{\text{cond}}(r) = \sum_{k} \left| \psi_{k}(r) \right|^2 f(E_{k} - \mu)$, where, for simplicity, we consider a single conduction band and suppress  the spin index. In a real AFM, the spin index should be kept, but it only 
modifies the amplitudes, not the position of the satellites. To describe 
the full crystal with $N$ unit cells at positions $R$, the density is 
replicated as $\rho(r) = \sum_{R} \rho_{\text{cond}}(r-R)$. This construction ensures that the Laue condition appears naturally, i.e., $F(q) = \int dr \, e^{i q r} \rho(r) 
     = S_0(q) \int_{\text{cell}} dr \, e^{i q r} \rho_{\text{cond}}(r)$, where $S_0(q) = \sum_R e^{iqR}$ is the lattice sum.  To evaluate $F(q)$, we need the explicit form of $|\psi_k(r)|^2$, namely
\begin{equation}
|\psi_k(r)|^2 = |\alpha_k|^2 |u_A(r)|^2 + |\beta_k|^2 |u_B(r)|^2 
+ \alpha_k \beta_k^* e^{i Q r} u_A(r) u_B^*(r) 
+ \alpha_k^* \beta_k e^{-i Q r} u_A^*(r) u_B(r).
\label{eqincom}
\end{equation}

Inserting Eq.~\ref{eqincom} into the diffraction amplitude $F(q)$, the 
Laue condition requires ordinary Bragg peaks at $q=G$, given by the first 
two terms of Eq.~\ref{eqincom}. The last two terms generate magnetic 
satellite peaks, satisfying
\begin{equation}
q \pm Q = G \quad \Rightarrow \quad 
q = G \pm Q = \frac{2 \pi m}{a} \pm \left(\frac{\pi}{a} + \delta\right).
\end{equation}
Thus, for $\delta=0$ one obtains the commensurate antiferromagnetic Bragg 
peaks at $q=G \pm \pi/a$, as observed in NiO and MnO, while for 
$\delta \neq 0$ the construction naturally produces incommensurate magnetic 
satellites at $q = G \pm (\pi/a + \delta)$, as seen in chromium. This result was demonstrated by Overhauser in 1962 using a magnetization approach\cite{Overhauser1962}. It is worth noting that, within our framework, the derivation follows in a more straightforward manner.

The last case involves a transition from metallic to insulating Mott behavior (MIT)\cite{MOTT1968}. In the metallic phase, the density of states near the Fermi level is high, and many electronic states satisfy \( f(E_{n\vec{k}} - \mu) \approx 1 \), meaning they are occupied and contribute significantly to the conduction electron density. As a result, the conduction-related structure factor \( F_{\text{cond}}(\vec{q}) \) is nonzero and reflects the presence of extended, delocalized electronic states. In contrast, the insulating phase is characterized by a finite energy gap \( \Delta > 0 \) around the Fermi level, such that no electronic states are available within this region. Accordingly, the Fermi function becomes sharply quantized: \( f(E_{n\vec{k}} - \mu) \approx 1 \) for states well below \( \mu \), and \( \approx 0 \) for those above. Near the Fermi level, occupations are frozen, suppressing the valence contribution to $F_{\text{cond}}$. Importantly, in both phases, the ionic structure factor \( F_{\text{ion}}(\vec{q}) \) remains essentially unchanged, as it is determined by the core electron distribution and the lattice geometry. Therefore, the observed changes in diffraction intensity across the transition originate solely from variations in \( F_{\text{cond}}(\vec{q}) \). This scenario represents a special case in which the conduction channel 
dominates and its contribution disappears in the insulating phase. 
However, depending on the material’s symmetry and the specific character 
of the Bloch states, the opposite trend $I_{\text{insulator}} > I_{\text{metal}}$ may arise, as demonstrated in resonant diffraction studies of correlated oxides\cite{Murakami1998,Wilkins2003}. An illustrative example is provided by $NaOsO_3$, where a forbidden Bragg reflection emerges across the Slater-type metal–insulator transition without structural distortions\cite{Calder2012}. We do not attempt to calculate this specific case here, but note that such behavior can be interpreted consistently within the present framework.
Thus, the detailed evolution of the intensity across the MIT provides a 
sensitive fingerprint of how conduction electrons reorganize in different 
electronic environments.

\section*{Conclusions}
We introduced the electronic structure factor $F_{\mathrm{cond}}$ as a 
phase-sensitive contribution to diffraction that encodes information about 
the occupied-band wave functions. Under Laue conditions $G=2\pi m/a$ with 
odd $m$ and with the ionic channel suppressed 
$\big(F_{\mathrm{ion}}(G)\!\approx\!0\big)$, the diffracted intensity is 
dominated by $I(G)\simeq \lvert F_{\mathrm{topo}}(G)\rvert^{2}$, which is 
governed by the relative sublattice phase $\Phi(k)$. The key ingredient is that $\Phi(k)$ enters only through $F_{\mathrm{cond}}$ and not through the ionic channel, turning diffraction into a genuinely phase-sensitive probe of electronic topology. In the SSH model, 
$\Phi(k)$ integrates to the Zak phase, allowing diffraction to 
discriminate between the trivial and topological regimes without any 
structural change.

This property makes $F_{\mathrm{cond}}$ a sensitive tool for 
characterizing topological phases and unconventional metal–insulator 
transitions. Moreover, $F_{\mathrm{cond}}$ can be responsive to applied 
magnetic or electric fields, since band dispersions and relative phases 
evolve under external control; this field tunability effectively turns 
diffraction into a spectroscopic probe of the electronic sector. In this 
sense, the measured intensity can convey substantially more information 
about the electronic state than the crystal structure alone.

In addition, the same Bloch-based construction naturally explains the 
appearance of magnetic Bragg satellites in antiferromagnets. For a 
commensurate modulation with ordering vector $Q=\pi/a$, additional peaks 
arise at $q=G\pm Q$, as observed in classical AFMs such as NiO and MnO. 
When the modulation becomes incommensurate, $Q=\pi/a+\delta$, the 
formalism yields displaced satellite peaks at 
$q=G\pm(\pi/a+\delta)$, in agreement with neutron diffraction results on 
chromium. This illustrates that the electronic contribution 
$F_{\mathrm{cond}}$ not only distinguishes topological phases in model 
systems such as SSH chains, but also captures the magnetic satellite 
structure of real antiferromagnets in a unified manner.

While our analysis used the ideal chiral-symmetric SSH limit as a 
transparent testbed, the formalism is general. Quantifying signal 
magnitudes with realistic atomic form factors, multiple scattering, 
finite temperature, disorder, and explicit chiral-symmetry breaking 
$(d_z\neq 0)$, as well as extending to higher-dimensional lattices and 
time-resolved experiments, are natural directions toward experimental 
deployment.

\section{acknowledgement}
SA acknowledges funding from DICYT regular 042431AP and Cedenna CIA250002.  D. G.-P. acknowledges ANID-Subdirección de Capital Humano/Doctorado Nacional/2023-21230818. This work is dedicated to the memory of J. d' Albuquerque e Castro, whose encouragement and scientific vision remain with us.

\appendix
\section{Chiral case: Berry connection and Zak phase}

For the occupied band in the chiral SSH model the Bloch eigenstate in the 
$\{A,B\}$ basis is
\begin{equation}
  \phi_A(k) = -\frac{e^{i\alpha(k)}}{\sqrt{2}}, 
  \qquad 
  \phi_B(k) = \frac{e^{i\beta(k)}}{\sqrt{2}},
\end{equation}
where in both sublattices carry equal weight  $|\phi_A|^2 = |\phi_B|^2 = \tfrac{1}{2}$. Then, the Berry connection is given by
\begin{equation}
  \mathcal{A}(k) = i\langle u_k|\partial_k u_k\rangle
                 = \tfrac{1}{2}\,\partial_k \alpha(k)
                 + \tfrac{1}{2}\,\partial_k \beta(k).
                 \label{conexberr}
\end{equation}
If we define the relative phase as $\Phi(k)=\beta(k)-\alpha(k)$, we can rewrite equation \ref{conexberr} as

\begin{equation}
  \mathcal{A}(k) = \partial_k \alpha(k) 
                 + \tfrac{1}{2}\,\partial_k \Phi(k).
\end{equation}

The first term $\partial_k \alpha(k)$ corresponds to a global phase that can be eliminated by choosing a convenient gauge. Therefore we have
\begin{equation}
  \mathcal{A}(k) = \tfrac{1}{2}\,\partial_k \Phi(k).
\end{equation}

Therefore, the Zak phase follows as
\begin{equation}
  \gamma = \int_{\mathrm{BZ}} \mathcal{A}(k)\,dk
         = \tfrac{1}{2}\,\Delta \Phi
         = \pi w \ \ (\mathrm{mod}\ 2\pi),
\end{equation}
with $w$ the winding number of $\Phi(k)$.

Finally, the intra-cell cross term entering the structure factor takes the form
\begin{equation}
  \phi_A(k)\,\phi_B^*(k) 
  = -\tfrac{1}{2}\,e^{-i\Phi(k)},
\end{equation}
so that the interference between the intracell geometry and the winding of 
$\Phi(k)$ discriminates the topological sectors $w=0$ and $w=1$.

\end{document}